\documentclass{iopart}

\usepackage{epsfig}
\usepackage{amssymb}

\begin{document}
\title[Kinetic path way of the Nematic-Isotropic phase transition]{Kinetic pathways of the Nematic-Isotropic phase
transition as studied by confocal microscopy on rod-like viruses.}
\author{M. Paul Lettinga\dag, Kyongok Kang\dag, Arnout Imhof\ddag, Didi Derks\ddag and Jan K. G. Dhont\dag}
\address{\dag
IFF, Institut Weiche Materie, Forschungszentrum J\"{u}lich,D-52425
J\"{u}lich, Germany }
\address{\ddag
Soft Condensed Matter, Debye Institute, Utrecht University,
Princetonplein 5, 3584 CC Utrecht, The Netherlands}

\ead{p.lettinga@fz-juelich.de}

%
%

\bibliographystyle{unsrt}

\maketitle

\newcommand{\hg}{\mbox{$h(\mathbf{r})$}}
\newcommand{\rr}{\mbox{$\mathbf{r}$}}
\newcommand{\kk}{\mbox{$\mathbf{k}$}}
\newcommand{\hkk}{\mbox{$\hat{\mathbf{k}}$}}
\newcommand{\qq}{\mbox{$\mathbf{q}$}}
\newcommand{\uu}{\mbox{$\hat{\mathbf{u}}$}}
\newcommand{\nn}{\mbox{$\hat{\mathbf{n}}$}}
\newcommand{\rot}{\mbox{$\hat{\mathcal{R}}$}}
\newcommand{\dg}{\mbox{$\dot{\gamma}$}}

\begin{abstract}
We investigate the kinetics of phase separation for a mixture of
rod-like viruses (\emph{fd}) and polymer (dextran), which
effectively constitutes a system of attractive rods. This
dispersion is quenched from a flow-induced fully nematic state
into the region where the nematic and the isotropic phase coexist.
We show experimental evidence that the kinetic pathway depends on
the overall concentration. When the quench is made at high
concentrations, the system is meta-stable and we observe typical
nucleation-and-growth. For quenches at low concentration the
system is unstable and the system undergoes a spinodal
decomposition. At intermediate concentrations we see the
transition between both demixing processes, where we locate the
spinodal point.
\end{abstract}


\section{Introduction}

Systems that are quenched into a state where at least one order
parameter is unstable undergo spinodal phase separation. Here, the
initially homogeneous system is unstable against fluctuations of
arbitrary small amplitude, and phase separation sets in
immediately after a quench. In the initial stage of phase
separation an interconnected "labyrinth structure" of regions with
somewhat higher and lower values of the order parameter is
observed. For quenches where the system becomes meta-stable, phase
separation is initiated by fluctuations with a sufficiently large
amplitude. Since such fluctuations have a small probability to
occur, phase separation sets in after a certain delay time,
referred to as the induction time. Here, nuclei are formed
throughout the volume which grow when they are sufficiently large.
The two different mechanisms of phase separation (spinodal
decomposition and nucleation-and-growth) can thus be distinguished
during the initial stages of phase separation from (i) the
difference in morphology (interconnected structures versus growth
of isolated nuclei) and (ii) the delay time before phase
separation sets in (no delay time for spinodal decomposition and a
finite induction time for nucleation-and-growth). As Onsager
showed in 1949\cite{Onsager49}, the situation is different when
the particles are not spherical in shape, i.e. disk-like or
elongated particles. Here the system can become unstable or
meta-stable with respect to fluctuations in \emph{orientation}.
These orientational fluctuations drive concentrations differences,
resulting in a phase with high concentration and orientational
order, the \emph{nematic} phase, and a phase with low
concentration and no orientational order, the \emph{isotropic}
phase. For very long and thin rods with short-ranged repulsive
interactions, the binodal concentrations, i.e. the concentrations
of the isotropic and nematic phases in equilibrium after phase
separation is completed, have been determined using different
approximations in minimizing Onsager's functional for the free
energy (see ref. \cite{Vroege92} and references therein), while
for shorter rods computer simulations have been performed to
obtain binodal concentrations \cite{Bolhuis97,Graf99}. Also the
spinodal concentration where the isotropic phase becomes unstable
has been found\cite{Onsager49,Kayser78}.

Binodal points are relatively easy to determine experimentally,
since they are given by the concentrations of the bottom and top
phase after phase separation. In contrast, it is not at all
straightforward to obtain spinodal points, since one would ideally
n like to perform a concentration quench from low or high
concentration into the two-phase region, where the initial state
is isotropic or nematic, respectively. In a recent paper such a
kind of 'quench' was performed by inducing polymerization of short
actin chains \cite{Viamontes05}, and tactoids and spinodal
structures were observed. Signatures of spinodal decomposition
have also been obtained for boehmite rods, by homogenizing a phase
separated system and sequential polarization microscopy and Small
Angle Light Scattering measurements\cite{vanbruggen99a}.
Alternatively, external fields like shear flow \cite{Lenstra01}
and a magnetic field \cite{Tang93,Lemaire05} can be applied to
prevent a system from phase separation and to stabilize the
nematic phase. After cessation of such an external field the
nematic phase will become unstable or meta-stable, depending on
the constitution of the sample, and phase separation sets in. In
this paper we induce a fully nematic phase with a well defined
director by imposing shear flow to a dispersion of colloidal rods.
We use \emph{fd}-viruses as system, since the equilibrium phase
behavior concerning the binodal points, has been well understood
on the basis of Onsager theory\cite{Fraden95,Chen93}. Polymer is
added to the dispersion in order to widen the region of
isotropic-nematic phase coexistence, which facilitates the phase
separation experiments \cite{Dogic04a}. We perform quenches of a
flow aligned initial state to zero shear, which renders the system
unstable or meta-stable to fluctuations in the orientation,
depending on the concentration of rods. As a consequence phase
separation sets in, which we observe by Confocal Laser Scanning
Microscopy (CSLM). We perform this experiment for different
concentrations, throughout the region of phase coexistence. Our
results illustrate the difference between nucleation-and-growth
and spinodal decomposition in the case of demixing elongated
particles, and result in the determination of the
nematic-isotropic spinodal point.

\begin{figure}
\begin{center}
\epsfig{file=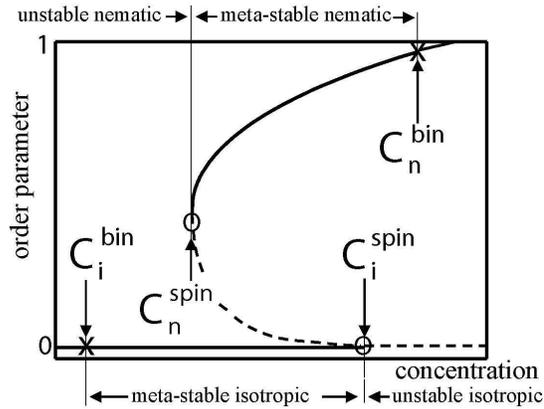,width=0.55\textwidth}
\caption{\label{stabil} The bifurcation diagram, where the
orientational order parameter $P_{2}$ is plotted against
concentration. Indicated are the various meta- or unstable regions
for the two different initial states of the homogeneous
suspension. The points marked by X and O are spinodal and binodal
points, respectively.}
\end{center}
\end{figure}

\section{On the Instability of Initial States}

A convenient way to analyze the stability of a homogeneous initial
state is to derive an equation of motion for the order parameter
tensor,
\begin{eqnarray} \label{ordertensor}
\mathbf{S}_{0}(t)\;=\;\oint d\uu\;\uu\uu\;P_{0}(\uu,t)\;.
\end{eqnarray}
The largest eigenvalue $P_{2}$ of the tensor
$\mathbf{Q}_{0}(t)=\mbox{\small{$\frac{3}{2}$}}\,[\,\mathbf{S}_{0}(t)
-\mbox{\small{$\frac{1}{3}$}}\,\hat{\mathbf{I}}\,]$ (where
$\hat{\mathbf{I}}$ is the identity) measures the degree of
alignment. For the isotropic state $P_{2}=0$, while for a
perfectly aligned state $P_{2}=1$. A stability analysis of
stationary solutions of this equation of motion is most
conveniently made on the basis of a bifurcation diagram
\cite{Kayser78}, where $P_{2}$ for stationary solutions is plotted
against the concentration. A schematic bifurcation diagram is
given in fig.\ref{stabil}. The two solid lines represent stable
stationary solutions of the equation of motion, while the dotted
lines represent unstable stationary solutions. The isotropic state
ceases to be stable above the concentration indicated as
$C_{i}^{spin}$, while the nematic state becomes unstable at
concentrations lower than $C_{n}^{spin}$. Above $C_{i}^{spin}$,
the isotropic state is still a stationary solution , but is now
unstable. Below $C^{spin}_{n}$, on the contrary, there is no
unstable nematic state that is a stationary solution of the
equation of motion. The two spinodal concentrations $C_{i}^{spin}$
and $C_{n}^{spin}$ are connected by a separatrix which separates
the basins of attraction for the isotropic and nematic state. A
homogeneous initial state above this separatrix develops a higher
degree of alignment, while an initial state below the separatrix
becomes more isotropic.

Note that the bifurcation diagram relates to homogeneous systems.
In an experiment, starting from a homogeneous state,
inhomogeneities develop simultaneously with the change of the
order parameter of the otherwise homogeneous system. In
equilibrium, after completion of phase separation, there is an
isotropic phase with concentration $C_{i}^{bin}$ in coexistence
with a nematic phase with concentration $C_{n}^{bin}$. One can
either start from a stationary state, in which case $P_{0}$ in
eq.(\ref{ordertensor}) is independent of time, or from a
non-stationary state, like a nematic state with a concentration
lower than $C_{n}$, in which case the time dependence of the
temporal evolution of alignment of the otherwise homogeneous
system couples to the evolution of inhomogeneities through the
time dependence of $P_{0}$.

In this paper we prepare an initial nematic state shearing a
suspension at large enough shear rate such that the induced
nematic phase is stable against phase separation (see
refs.\cite{Dhont03c} for a discussion of the bifurcation diagram
for sheared systems), and then quench to zero shear rate. For this
initial state it is expected that spinodal decomposition occurs at
lower concentrations, while nucleation and growth is observed at
higher concentrations. For an isotropic initial state this would
be reversed : spinodal decomposition at high concentrations and
nucleation and growth at lower concentrations. The observed phase
separation kinetics thus depends crucially on the preparation of
the initial state of the suspension.

\section{Materials and methods}\label{Materials and methods}

As model colloidal rods we use \emph{fd}-virus particles which
were grown as described elsewhere\cite{Dogic04a}. A homogeneous
solution of $22.0 \;  mg/mL$ {\it fd}-virus and $12.1 \; mg/mL$ of
Dextran (507 kd, Sigma-Aldrich) in $20 mM$ tris buffer at
$pH\;8.15$ with $100 mM$ NaCl is allowed to macroscopically phase
separate. This concentration of {\it fd}-virus is exactly in the
biphasic region, which is very small when no polymer is added,
namely between $21 $ and $23 \; mg/mL$. Due to the added polymer,
the binodal points shift to $17$ and $30\;  mg/mL$, respectively.
New dispersions are prepared by mixing a known volume of the
coexisting isotropic and nematic bulk phases. The relative volume
of nematic phase in this new dispersion is denoted as
$\varphi_{nem}$.

For the microscopic observations we used a home-built counter
rotating cone-plate shear cell, placed on top of a Leica TCS-SP2
inverted confocal microscope. This cell has a plane of zero
velocity in which objects remain stationary with respect to the
microscope while shearing. For details of the setup we refer to
reference\cite{Derks04}. For the measurements described here we
used confocal reflection mode at a wavelength of 488 nm. Quench
experiments were done as follows. Samples were first sheared at a
high rate of $10\; s^{-1}$ for several minutes. The shear was then
suddenly stopped, after which images were recorded at regular time
intervals. These images were parallel to the flow-vorticity plane.
The table gives an overview of the concentrations where quench
experiments have been performed.

\begin{table}
\caption{Overview of the used samples.}
\begin{tabular}[h]{|c|c|c|c|c|c|c|c|}
  \hline
   Code &  $\varphi_{nem}^5$ & $\varphi_{nem}^4$ & $\varphi_{nem}^3$ & $\varphi_{nem}^2$ & $\varphi_{nem}^1$  \\ \hline
  \emph{fd} ($mg/mL$) & 29.5 & 28.1 & 25.8 & 23.6 & 19.3 \\ \hline
  $\varphi_{nem}$& 0.96 & 0.85 & 0.68 & 0.52 & 0.18 \\ \hline
\end{tabular}

\end{table}

\begin{figure}
\begin{center}
\includegraphics[width=.95\textwidth]{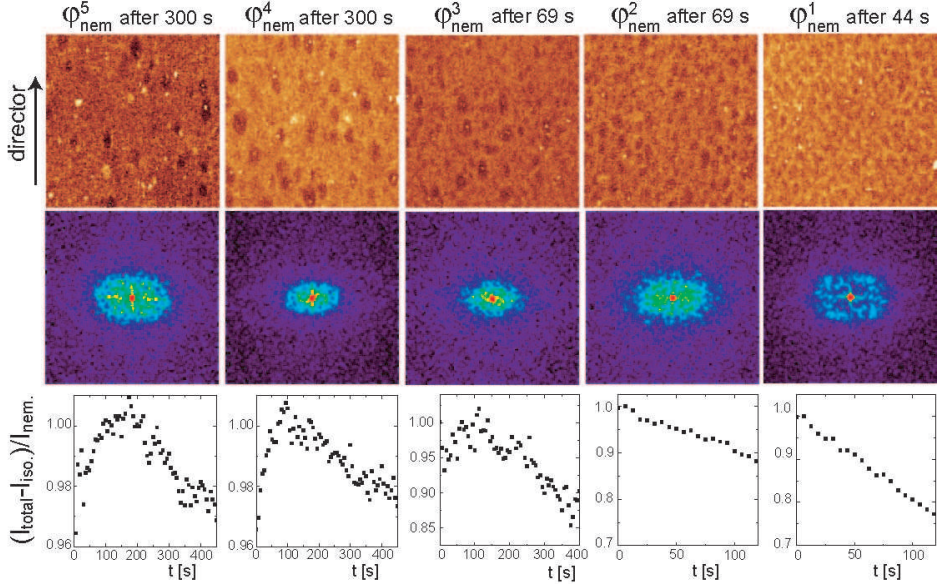}
\caption{\label{initial} The initial stages of phase separation
for five different concentrations after a quench from a flow
aligned nematic phase to zero shear. The top row shows the
micrographs taken by reflection confocal scanning laser microscopy
(field of view $=110\;\mu m$); the middle row shows the fourier
transform of the micrographs; the bottom row plots the mean
intensity of the micrographs minus the mean intensity for the
isotropic phase, normalized by the initial intensity of the
nematic phase.}
\end{center}
\end{figure}

\section{Results}

In the top row of Figure \ref{initial} we show micrographs of the
initial stage of phase separation for five different
concentrations taken after a shear rate quench from a high shear
rate, where the nematic state is stable for each concentration, to
zero shear. These images show the flow (vertical) - vorticity
(horizontal) plane at a given time after the quench. Thus the
director of the initial nematic phase lies in the vertical
direction. Fourier transforms of the images are plotted in the
second row of Fig. \ref{initial}. The background is corrected for
by subtracting the Fourier transform of the first frame. The third
row plots the development of the total intensity of the images
minus the intensity in the isotropic phase, as determined from an
isolated isotropic region, normalized by the initial nematic
intensity. Qualitatively the difference between the concentrations
is  obvious. In the first two images, i.e. the two highest
concentrations, isolated dark ellipsoidal structures can be seen
on a bright back ground. These are droplets of the isotropic phase
referred to as tactoids. The number of tactoids increases when the
concentration is decreased (b and c) until the structures become
interconnected (d and e). This also follows from the fourier
transform of the pictures where a ring is detected for the lowest
concentration and a constant increasing intensity towards K=0 for
the highest concentration. The time scale at which the
inhomogeneities are formed also changes. As can be seen in the
third row of Fig. \ref{initial}, the high concentrations all show
an induction time before the phase separation sets in, while for
the low concentrations phase separation sets in immediately. Note
also the times at which the images in Fig. \ref{initial} were
taken. The isolated nuclei and the induction time are typical for
nucleation-and-growth, while the interconnected structures and
immediate phase separation are typical for spinodal decomposition.

\begin{figure}
\begin{center}\includegraphics[width=.65\textwidth]{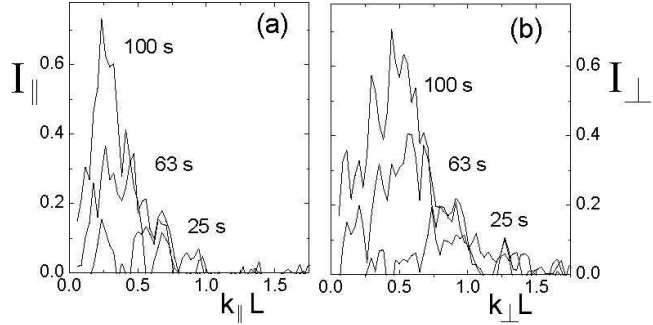}
\caption{\label{FFT_profile} The cross section of the Fourier
transform parallel (a) and perpendicular (b) to the director for
 $\varphi_{nem}^1$.  The length is scaled by the rod length $L$.}
\end{center}
\end{figure}

\begin{figure}
\begin{center}
\includegraphics[width=.75\textwidth]{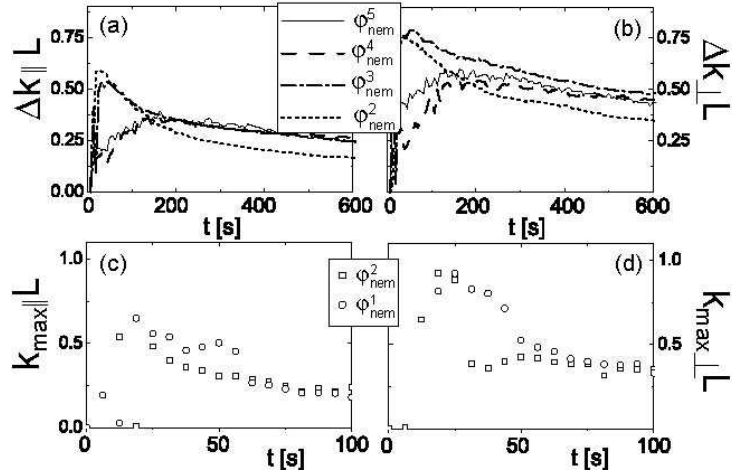}
\caption{\label{qmax_qwidth}  Width of the 2-D gaussian fit,
$\Delta k L$ , of the Fourier transform parallel (a) and
perpendicular (b) to the director for the higher concentrations
for the higher concentrations. Wavevector $k_{max}L$ where the
intensity is maximum for the cross sections parallel (c) and
perpendicular (b) to the director for the lower concentrations. }
\end{center}
\end{figure}

\begin{figure}
\begin{center}
\includegraphics[width=.75\textwidth]{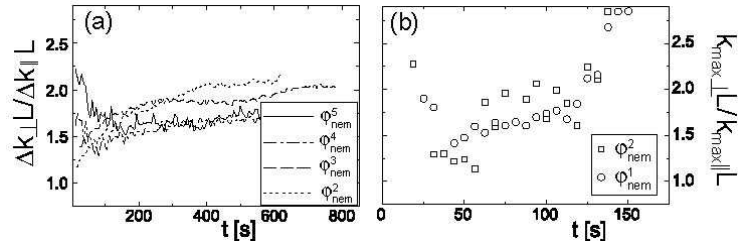}
\caption{\label{ratio} (a) The ratio $ \Delta k_{\bot} L / \Delta
k_{\|} L$ for the higher concentrations  and (b) the ratio
$k_{max,\bot} L /k_{max,\|} L$ for the lower concentrations.}
\end{center}
\end{figure}

We use the Fourier transform of the images as shown in Fiq.
\ref{initial} to quantify the phase separation processes. The
interesting quantity for nucleation-and-growth is the width of the
Fourier transform, $\Delta k$, which is a measure for the
anisotropic form factor of the nuclei. Alternatively one could
determine the average size of the features in real space, but due
to the low contrast this is difficult. For spinodal decomposition
the interesting quantity is the wavevector at which the fourier
transform reaches its maximum, $k_{max}$, quantifying the fastest
growing concentration fluctuation. In both cases the fit of the
Fourier transform should be performed in two dimensions, since the
initial state is anisotropic. Therefore we took cross sections
parallel and perpendicular to the director in the Fourier domain,
i.e. the vertical and horizontal in Fig. \ref{initial} middle row,
to  determine $k_{max}$. Typical cross sections for are shown in
Fig. \ref{FFT_profile}, where the wavevector $k$ is scaled by the
rod length $L$. To determine $\Delta k$, we performed a 2-D
gaussian fit around the origin of the Fourier transforms. Results
of a 2-D gaussian fit of the Fourier transform around the origin
are shown for the higher concentrations in Fig. \ref{qmax_qwidth}a
and b, plotting the width in the direction of the director and
perpendicular to the director, respectively. $k_{max}$ as found
from fits of the cross sections parallel and perpendicular to the
director are given in Fig. \ref{qmax_qwidth} c and d,
respectively. Both fit procedures result in an anisotropic
morphology as can be seen in Fig. \ref{ratio}, where we plotted  $
\Delta k_{\bot} L / \Delta k_{\|} L$ and
$k_{max,\bot}L/k_{max,\|}L$.

\begin{figure}
\begin{center}

\includegraphics[width=.75\textwidth]{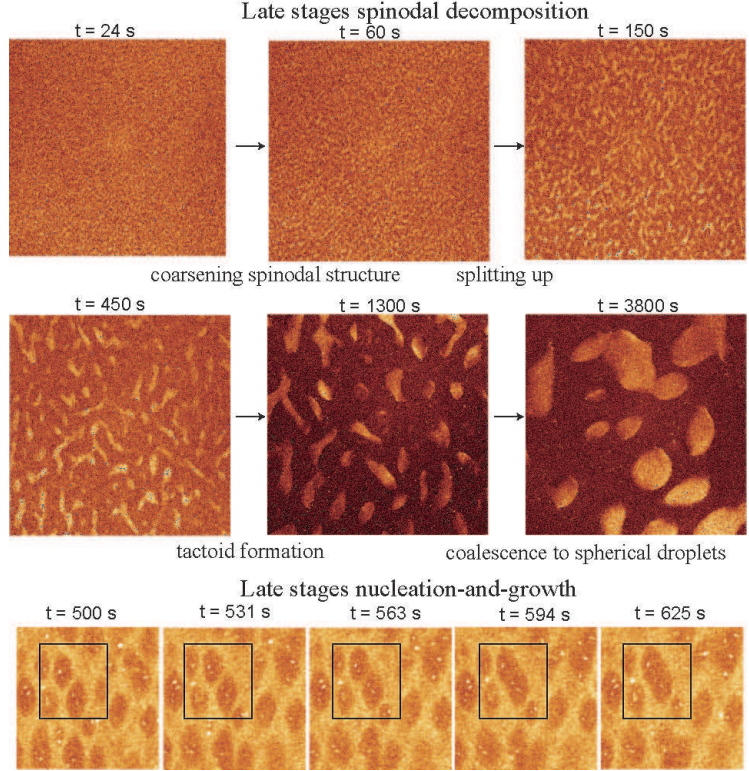}
\caption{\label{latestage} The late stages for spinodal
decomposition in the top two rows ($\varphi_{nem}^1$, field of
view=$375 \;\mu m$), and coalescence of tactoids in the bottom row
($\varphi_{nem}^5$, field of view=$73 \;\mu m$). }
\end{center}
\end{figure}

The late stages of the different phase separation processes also
show some interesting phenomenology, as can be seen in Fig.
\ref{latestage}. For spinodal decomposition we observe first a
growing of the interconnected structures, which then break down
into tactoids. Later on tactoids coalesce, and they become more
spherical with increasing size. Note that these tactoids contain
the nematic phase and not the isotropic phase, as observed for the
nucleation-and-growth process at higher concentrations. In the
late stage of nucleation-and- growth, i.e. at high concentrations,
we see that coalescence of tactoids containing the isotropic phase
as shown in the bottom row of Fig. \ref{latestage} is favorable
when two tactoids meet somewhat from the middle. In this case the
rod orientation near both features is similar and the barrier
which has to be overcome for coalescence is low.

\section{Discussion}

On the basis of these observations we can now locate the
metastable region, i.e. where the system has to overcome a free
energy barrier, and the unstable region, where there is no such
barrier. At the high concentrations ($\varphi_{nem}^5$,
$\varphi_{nem}^4$) the system is meta-stable, which is reflected
by the observed isolated structures formed (top row in
Fig.\ref{initial}) and the induction time (bottom row in
Fig.\ref{initial}). With decreasing concentration the system
approaches  the unstable region: the number of nuclei increases
while the induction time decreases and finally vanishes. The
lowest concentration $\varphi_{nem}^1$ is clearly unstable after
cessation of the flow. It shows all the features typical for
spinodal decomposition: phase separation immediately sets in
throughout the whole sample, with a typical length scale which is
characterized by the scattering ring observed in the Fourier
transform. It can be shown, in fact, that the observed phase
separation process for the lowest concentration has features
typical for the spinodal decomposition of rods, as derived
recently from a microscopic theory by one of the
authors\cite{Dhont05}. This will be the subject of a following
paper\cite{Lettinga05d}.

In the intermediate region it is difficult to judge from the
morphology if nucleation-and-growth takes place or spinodal
decomposition, since it is difficult to distinguish between a high
number of tactoids and an interconnected structure. However,
$\varphi_{nem}^3$ shows a short induction time after the quench
after which clearly separated tactoids are formed, while for
$\varphi_{nem}^2$ phase separation immediately sets in showing
ellipsoidal structures which clearly 'influence' each other.
Moreover, Figures \ref{qmax_qwidth} shows that the size of the
structures formed in $\varphi_{nem}^3$ coincides after some time
with the clearly nucleated structures of $\varphi_{nem}^4$ and
$\varphi_{nem}^5$, while the size of the structures formed in
$\varphi_{nem}^2$ coincides with samples which clearly show
spinodal decomposition. Thus, we locate the transition from
meta-stable to unstable, i.e. the spinodal point, between at $23.5
\; mg/mL$ and $25.8 \; mg/mL$. This is the first experimental
observation of the spinodal point in a rod-like system. We should
mention at this point that in fact our sample consists of a
mixture of rods and polymer. Addition of the polymer causes a
widening of the biphasic region \cite{Dogic04a}, i.e. a shift of
the binodal points. It is now interesting to see that the high
concentration binodal shifts as much as from $23 \; mg/mL$ to $30
\; mg/mL$. In contrast, the high concentration binodal point,
$C^{bin}_n$, shifts from a concentration between $21\;mg/mL$ and
$23\;mg/mL$ to somewhere between $23.5 \; mg/mL$ and $25.8 \;
mg/mL$. This leads to the interesting conclusion that the shift of
the high concentration binodal point, $C^{bin}_n$, due to the
attraction between the rods, is considerable compared to the shift
of the high concentration spinodal point, $C^{spin}_n$. In other
words, making the rods attractive causes a widening of the
meta-stable region, while the unstable region remains unaffected.
Addition of more polymer will result in more complex kinetics as
described in reference \cite{Dogic03}.

Interestingly, for all concentrations we observe that the
morphology of the phase separating system is anisotropic. This is
most clear for the highest concentrations, where the tactoids all
point upwards, i.e. in the direction of the director of the
surrounding nematic phase. Also the Fourier transforms for the
lower concentrations show deformed intensity rings in fourier
space (most right FFT image in Fig. \ref{initial}). Moreover, also
the kinetics of the phase separation is fastest in the direction
of the nematic director. This follows for instance from the ratio
of $\mathbf{k_{max}}$ as plotted in \ref{ratio}b, which increases
in time. In other words, for all concentrations phase separation
is anisotropic, due to residual alignment after the quench of the
initially strongly sheared suspension, and not isotropic as is the
case for spheres\cite{Aarts04}.

The length of the first observed tactoids just below $C^{bin}_n$
is about 12 times the rod length, while just above $C^{spin}_n$ it
is seven times the rod length. The thickness is about two third of
the length in both cases. Typical length scales for the initial
spinodal morphology are not more than six rod lengths. These sizes
seem to be quite small, considering also the random orientation of
the rods in the isotropic phase, but it is in accordance with a
the microscopic theory for spinodal decomposition of
rods\cite{Dhont05}. It does suggest that we really image the
initial stage. The breaking up of the spinodal structure into
nematic tactoids and the sequential growth in the late stage of
spinodal decomposition seems surprising since for dispersions of
spheres only coalescence and macroscopic phase separation would be
observed. However, a similar order of events has been observed for
polymer mixtures with thermotropic liquid crystals\cite{Nakai96}.
Simulations on such mixtures show that the break down is due to
the effect of the flow-alignment coupling, and not primarily due
to elastic effects\cite{Araki04}. An explanation in the same line
was given by Fukuda in a numerical treatment of time-dependent
Ginzburg-Landau equations of liquid crystalline polymers
\cite{Fukuda99}. The volume dependence of the morphology in the
final stage can be explained by the competition between the
interfacial tension and nematic elasticity of the
tactoids\cite{Prinsen03}.

\section{Conclusion}

We studied the kinetics of the nematic-isotropic phase transition
of a dispersion of \emph{fd}-virus particles with added polymer
after shear quenches into the two-phase region. By varying the
equilibrium rod concentration $\varphi_{nem}$ we were able to
detect a nucleation-and-growth mechanism for high $\varphi_{nem}$,
spinodal decomposition for low  $\varphi_{nem}$, and the
transition between the two processes. In this way we were able to
trace for the first time the nematic-isotropic spinodal point
$C^{spin}_n$. Thus, we found that addition of polymer widens the
meta-stable region greatly. Furthermore, we showed that the phase
separation is strongly influenced by the director of the initial
nematic state. The nematic phase also influences the late stages
of spinodal decomposition, causing a splitting up of the
interconnected structures.

\ack This work was performed within the framework of the
Transregio SFB TR6, "Physics of colloidal dispersions in external
fields".

\end{document}